\documentclass[twocolumn,english,pra]{revtex4}
\usepackage[T1]{fontenc}
\usepackage[latin9]{inputenc}
\usepackage{bm}
\usepackage{amsmath}
\usepackage{amssymb}
\usepackage{graphicx}
\usepackage{esint}

\makeatletter
\@ifundefined{textcolor}{}
{%
 \definecolor{BLACK}{gray}{0}
 \definecolor{WHITE}{gray}{1}
 \definecolor{RED}{rgb}{1,0,0}
 \definecolor{GREEN}{rgb}{0,1,0}
 \definecolor{BLUE}{rgb}{0,0,1}
 \definecolor{CYAN}{cmyk}{1,0,0,0}
 \definecolor{MAGENTA}{cmyk}{0,1,0,0}
 \definecolor{YELLOW}{cmyk}{0,0,1,0}
 }

\@ifundefined{definecolor}
 {\usepackage{color}}{}

\usepackage{babel}

\makeatother

\usepackage{babel}
\begin{document}

\title{Universality of the Berezinskii-Kosterlitz-Thouless type of phase
transition \\
 in the dipolar XY-model}

\author{A.\,Yu.\,\,Vasiliev$^{1,2}$, A.\,E. Tarkhov$^{1,3}$, L.\,I.
Menshikov$^{1,4}$, P.\,O. Fedichev$^{1,2}$, and Uwe R. Fischer$^{5}$}

\affiliation{$^{1}$Quantum Pharmaceuticals Ltd, Ul.\,\,Kosmonavta Volkova 6-606,
Moscow, Russia}

\affiliation{$^{2}$Moscow Institute of Physics and Technology, Institutskii per.\,9,
Dolgoprudny, Moscow Region, 141700, Russia}

\affiliation{$^{3}$Physics Department, M.V. Lomonosov Moscow State University,
Vorobievy gory, Moscow, 119992, Russia}

\affiliation{$^{4}$Northern (Arctic) Federal University, Severnaya Dvina Emb.
17, Arkhangelsk, 163002, Russia}

\affiliation{$^{5}$Seoul National University, Department of Physics and Astronomy,
Center for Theoretical Physics, 151-747 Seoul, Korea}
\begin{abstract}
We investigate the nature of the phase transition occurring in a planar
XY-model spin system with dipole-dipole interactions. It is demonstrated
that a Berezinskii-Kosterlitz-Thouless (BKT) type of phase transition
always takes place at a finite temperature separating the ordered
(ferro) and the disordered (para) phases. The low-temperature phase
corresponds to an ordered state with thermal fluctuations, composed
of a {}``gas'' of bound vortex-antivortex pairs, which would, when
considered isolated, be characterized by a constant vortex-antivortex
attraction force which is due to the dipolar interaction term in the
Hamiltonian. Using a topological charge model, we show that small
bound pairs are easily polarized, and screen the vortex-antivortex
interaction in sufficiently large pairs. Screening changes the linear
attraction potential of vortices to a logarithmic one, and leads to
the familiar pair dissociation mechanism of the BKT type phase transition.
The topological charge model is confirmed by numerical simulations,
in which we demonstrate that the transition temperature slightly increases
when compared with the BKT result for short-range interactions. 
\end{abstract}
\maketitle

\section{Introduction}
The standard paradigm of phase transitions in a planar system of electrically
neutral particles representing an effective XY-model spin dictates
that long-range order does not exist at any finite temperature \cite{Mermin,Hohenberg}.
On the other hand, below a critical temperature $T_{c}$, spin-spin
correlations in two spatial dimensions decay in a power-law fashion.
As a consequence, short-range cooperative phenomena, such as superfluidity,
can exist at these low temperatures. Below the critical point, thermal
excitations predominantly occur in the form of vortex-antivortex pairs.
Due to the attraction between vortex and antivortex, the vortex-antivortex
pairs remain bound at $T<T_{c}$, and the gas thus remains superfluid,
while the dissociation of vortex-antivortex pairs at $T>T_{c}$ leads
to an exponential decay of correlations. This standard scenario, corresponding
to the celebrated BKT phase transition \cite{Berezinskii,kosterlitz1973oma,kosterlitz1974critical},
however conventionally only applies when the particle interact by
short-range (contact) potentials.

Spin ordering phenomena in planar systems with long-range forces mediated
by the dipole-dipole interaction between particles carrying static
dipole moments $\mu$, have been extensively studied in the past,
cf., e.g., \cite{maleev1976dft,pokrovsky1977zhetf,Feigelman1979,Yafet,Seul,kashuba1993stripe,kaplan1993domain,Beloshapkin,rapini2007phase,mol2011phase,weis2003simulation,MaierSchwabl,Bune,deBell2000dipolar,Vedmedenko2000,Rastelli,giuliani2007striped,Vaz  Bland and Lauhoff,baek2011}.
The dipole-dipole interaction in many cases tends to stabilize the long-range
order against thermal fluctuations, and the ground state of the spin
system may thus be spontaneously polarized \cite{maleev1976dft,pokrovsky1977zhetf,Feigelman1979,Yafet,MaierSchwabl},
or acquire various structures \cite{Seul,kashuba1993stripe,kaplan1993domain,Beloshapkin,rapini2007phase,mol2011phase,Bune,deBell2000dipolar,Vedmedenko2000,Rastelli,weis2003simulation,giuliani2007striped,Vaz  Bland and Lauhoff}.
So far, most efforts aimed at understanding a possible phase transition
in such systems were a combination of renormalization group (RG) arguments
with phenomenological approaches \cite{pokrovsky1977zhetf,Feigelman1979,Yafet,Seul,kashuba1993stripe,kaplan1993domain,MaierSchwabl,deBell2000dipolar,giuliani2007striped},
or Monte Carlo and Molecular Dynamics simulations \cite{Beloshapkin,rapini2007phase,mol2011phase,Vedmedenko2000,Rastelli,weis2003simulation}.
However, it is fair to say that a complete picture of the nature of
the phase transition {in the particular case of the two-dimensional
(2D) dipolar XY model, where the dipoles sample the {\em full anisotropy}
of the dipolar interaction, is elusive. While Refs.\,\cite{rapini2007phase,mol2011phase}
state that a BKT type transition is observed from their numerical
results, a physical mechanism explaining the phase transition is missing.
To clarify the nature of this transition is the main aim of our work.

{Possible applications, once a thorough understanding of the 2D dipolar
XY model has been obtained, span a broad range \cite{weis2003simulation},
of which we quote just a few. Besides the commonly studied ferromagnetic
and ferroelectric thin films (cf., e.g., \cite{deBell2000dipolar,Bune}),
these include ultracold dilute gases. There are currently major efforts
undertaken to cool heteronuclear molecules with large dipole moments
to quantum degeneracy \cite{Zirbel,Chotia,Carr}, which will ultimately
lead to studies of the dipolar BKT transition \cite{Prokofiev,Dalibard,Chin,Baranov}.
Effective spin models with dipolar interactions in the highly controllable
environment of ion traps have received considerable attention as well
\cite{Blatt,Cirac}. In a biological context, dipole-dipole interactions
determine, for example, the formation of a 2D hydrogen-bond network
and the large-scale polarization of water molecules in the hydration
layers of proteins \cite{Oleinikova,Matyushov}. 
Finally, we note
that {\em confinement} in isolated vortex-antivortex pairs by the
string tension is one of the rare instances outside the realm of Quantum
Chromodynamics, in which linear interaction potentials between a particle
and its antiparticle, in Quantum Chromodynamics between quark and
antiquark, occur \cite{QCDreview,Adams}.

For contact interactions, vortex and antivortex in a vortex-antivortex
pair interact by a logarithmic potential, which in superfluids is
due to kinetic energy of the flow, and the attraction force decays
with the inverse distance. This logarithmic interaction is the primary
requirement for the occurrence of the BKT transition. On the other
hand, in the presence of a dipolar interaction term in the Hamiltonian,
a spatially constant attraction force $K_{0}$ between vortex and
antivortex in an isolated pair occurs \cite{Feigelman1979,MaierSchwabl} (also see below),
which bears an obvious potential importance for the phase transition,
which has been overlooked in most previous investigations of the 2D
dipolar XY model, with the notable exception of Maier and Schwabl \cite{MaierSchwabl}.

The latter detailed consideration of the 2D dipolar XY-model within a RG treatment has shown, quoting \cite{MaierSchwabl},
that the {}``...flow diagram of the ferromagnetic transition is strikingly
similar to the Kosterlitz-Thouless transition.'' 
On the other hand, the existence of the linear interaction between vortices, the vortex
confinement, led the authors of \cite{MaierSchwabl} to the statement
that a novel phase transition, distinct from BKT, takes place in the
2D dipolar XY model. In other words, the dipolar interaction is argued there to be relevant for the nature 
of the phase transition.
In the following, we critically examine the latter conclusion. We demonstrate
that the sole effect of the dipole-dipole interaction is that a vortex-antivortex
pair dissociation transition of the BKT type occurs at slightly higher
temperatures. To this end, we use an analytical model, applicable
sufficiently close to the transition point, backed up by numerical
simulations for the full range of temperatures. We show that the apparently
inconsistent pictures of confinement of isolated vortex-antivortex
pairs and occurrence of a phase transition driven by the familiar
BKT mechanism of pair dissociation can be made fully consistent with
each other if one correctly accounts for shielding of the bare vortex-antivortex
pair tension at finite temperatures. The shielding effect of the linear
attraction potential within a large pair is mediated by the large
number of small vortex-antivortex pairs in which it is immersed. The
vortex-antivortex interaction only remains linear at small distances
$R$ between the vortices in a pair, $R<r_{0}$, where the length
parameter $r_{0}$ is defined in Eq.\,(\ref{eq: scale r_0}) below and discussed in more detail 
in the Appendix \ref{r0}; it however becomes logarithmic at sufficiently large vortex-antivortex
pair size, $R>r_{0}$. Hence, we conclusively demonstrate that the dipolar interaction contribution is irrelevant 
in the sense of the RG.

\section{Vortex Free Energy in the dipolar XY model}  
The polarization (effective spin) states of the 2D dipolar XY model,
which correspond to a vortex-antivortex pair gas, are described by
the continuous two-component vector-field $\bm{s}(\bm{r})=s\cos\theta{\bm{e}}_{x}+s\sin\theta{\bm{e}}_{y}$,
representing the hydrodynamic, coarse-grained average of the spin,
where $\theta$ is azimuthal angle. 
The topological charge inside a 2D contour C is defined as usual to be
$Q = (2\pi)^{-1} \int_{\cal C}d\theta$. It is the number 
of rotations of vector ${\bm s}$ ``winding'' around the vortex core along
closed contour $\cal C$; $Q\neq 0$ implies that some vortices
are inside $\cal C$. 
One can locate the vortex core by contracting the contour to a point. If
$Q \neq 0$ is retained in the process, the vortex core resides
at this point. For example, the upper vortex in
Fig.1 (left) has $Q = 1$ and the lower one $Q = -1$.

Note that the spin vortices are {\em dual} to the conventional superfluid vortices
in the sense that the spin vector is always oriented perpendicular to the corresponding ``fluid'' flow direction,  cf.\,Fig.\,\ref{energy density}. When the wave function of the superfluid is written in Madelung
representation $\Psi=\sqrt n \exp[i\theta]$, for constant density $n$ we have the kinetic energy $\propto \int df (\nabla\theta)^2$ 
(where $df$ is the surface element). For spin vortices, the kinetic energy reads 
$\propto \int df \sum _{a,b}\nabla_a s_b \nabla_a s_b$, see Eq.\,\eqref{eq: G H} below, which    
transforms for constant $s$ into an identical expression, $\propto \int df (\nabla\theta)^2$.  

In dimensionless form, the hydrodynamic
free energy functional (playing the role of the hydrodynamic ``Hamiltonian'' of the system) is 
\begin{equation}
G_{S}\left[\bm{s}(\bm{r})\right]=G_{0}+G_{dd}\label{eq: Hamiltonian}
\end{equation}
The short-ranged term $G_{0}$ generally reads ($a,b=x,y$) \cite{LL8}:
\begin{equation}
G_{0}=\frac{1}{2}\int df[C\sum_{a,b}\nabla_{a}s_{b}\nabla_{a}s_{b}+C'(\nabla\cdot\bm{s})^{2}+g\left(s^{2}-1\right)^{2}], 
\label{eq: G H}
\end{equation}
where $C,C'$ are constants, and $g>0$
is a {}``contact interaction'' coupling.
The {}``elastic'' constants $C,C'$ determine the ground-state spin
structure, which for $C>C'\ge0$ becomes ferromagnetic. For concreteness,
we set $C\equiv1$ as the unit of energy, and $C'=0$, as well as
put $g\equiv1$, choosing as our unit of length the vortex core size
(an ultraviolet cutoff). This does not affect our results on the nature
of the phase transition [also see the remark after Eq.\,\eqref{eq: Simplified VAP interaction}].
Note that in the hydrodynamic expression \eqref{eq: Simplified VAP interaction},
the constants $C,C'$ in \eqref{eq: G H} only affect the ultraviolet
cutoff, which is irrelevant for the nature of the phase transition;
the same applies to the coupling $g$ in \eqref{eq: G H}. 

The dipolar interaction
energy functional entering the hydrodynamic free energy in Eq.\,\eqref{eq: Hamiltonian} 
is given by \cite{maleev1976dft,pokrovsky1977zhetf,MaierSchwabl}
\begin{equation}
G_{dd}=\Lambda\int dfdf^{\prime}\frac{\rho_{P}\left(\bm{r}\right)\rho_{P}\left(\bm{r^{\prime}}\right)}{|\boldsymbol{r}-\boldsymbol{r}^{\prime}|}.\label{eq: dipole-dipole interaction}
\end{equation}
Here $\rho_{P}\left(\bm{r}\right)=-\nabla\cdot\bm{s}\left(\bm{r}\right)$
is the density of polarization charges, and $\Lambda\propto\mu^{2}$
represents the dipole-dipole interaction coupling constant. 
In Appendix \ref{microscopic}, we provide some considerations on the microscopic derivation of the above 
effective Hamiltonian \eqref{eq: Hamiltonian}.

We first undertake a qualitative consideration of the scaling behavior of the contributions 
to the hydrodynamic free energy functional of the planar 
dipolar XY model for a single vortex-antivoftex pair. 
Note that as in superfluids (see \cite{LLSP}, for example), the en-
ergy $G_0$ increases approximately proportional to $Q^2$. Therefore
the existence of vortices with $|Q|$ > 1 at low tem-
peratures is energetically disfavored.
Focusing our attention on the regime $T\approx T_{c}$, we have, typically,
that the vortex-antivortex pair size $R\gg1$. For these vortex-antivortex
pairs, using that $\nabla\cdot\bm{s}\sim1/R$ and hence $G_{0}\sim\log R$,
$G_{dd}\sim R$. Therefore, the energy of a vortex-antivortex pair
with $R\gg1$ increases linearly with their size, and vortex and antivortex
attract each other with a constant force \cite{Feigelman1979,MaierSchwabl},
in sharp distinction to contact interactions, 
where the force decreases as the inverse of the vortex-antivortex pair size.

For contact interactions, it is well established that the total free energy of the system can be written in the form 
of a double sum ($i,j=1\ldots2N$,
where $N$ is the number of vortex-antivortex pairs)
\begin{equation}
G_{S}=-\frac{1}{2}\sum_{i,\, j}q_{i}q_{j}u\left(r_{ij}\right)=\frac{1}{2}\sum_{i}q_{i}\Phi\left(\bm{r}_{i}\right).\label{eq: Simplified VAP interaction}
\end{equation}
where, in the {\it purely contact interaction} case, $u\left(r\right)=2\pi\log\left(1+\alpha r\right)$, with $\alpha$ is a constant, 
and $q_{i}=\pm1$ are the topological charges of the vortices \cite{kosterlitz1973oma,kosterlitz1974critical}
located at $\bm{r}_{j}$=$\left(x_{j},y_{j}\right)$, $r_{ij}=\left|\bm{r}_{i}-\bm{r}_{j}\right|$.

On the other hand, when strong dipolar interactions come into play,
which lead to {\em long-ranged and anisotropic} forces in the planar dipolar XY model, the validity of the pairwise summation formula \eqref{eq: Simplified VAP interaction} is, in distinction to the 
contact interaction case, highly nontrivial and needs to be thoroughly justified. Because Eq.\,\eqref{eq: Simplified VAP interaction} is at the heart of our analytical description in terms of plasma physics (see below), we have therefore set up the corresponding Langevin Dynamics simulations and ran
extensive checks to establish the validity of Eq.\,\eqref{eq: Simplified VAP interaction}
(see for a detailed discussion of the numerical procedure the Appendix \ref{Numerics}). 
For an illustration, typical results for the polarization field 
and energy distributions are presented in Fig.\,\ref{energy density}.

We find that the function 
\begin{equation}
 u\left(r\right)=2\pi\log\left(1+\alpha r\right)+K_{0}r
\end{equation}  
describes very well the energy of a vortex-antivortex pair system in the dipolar gas \cite{Linear behavior at small r}.
The quantity $K_{0}\equiv K_{0}\left(\Lambda\right)$
is the ``vortex-antivortex pair tension'' coefficient, depending
on the coupling strength $\Lambda$. The logarithmic contribution
in $u\left(r\right)$ arises from $G_{0}$ in Eq.\,(\ref{eq: G H})
and describes the interaction of vortices in the limit of $\Lambda\rightarrow0$
\cite{Berezinskii,kosterlitz1973oma,kosterlitz1974critical}.
The same numerical calculations also let us establish the functional
dependence of $K_{0}(\Lambda)$, which is presented in the inset to Fig.\,\ref{fig:Computed-transition-temperature} below.  
We observe a saturation in the increase of $K_0$ with $\Lambda$, an effect explained in detail in Appendix \ref{Numerics}.


\begin{figure}[t]
 \includegraphics[width=0.4933\columnwidth]{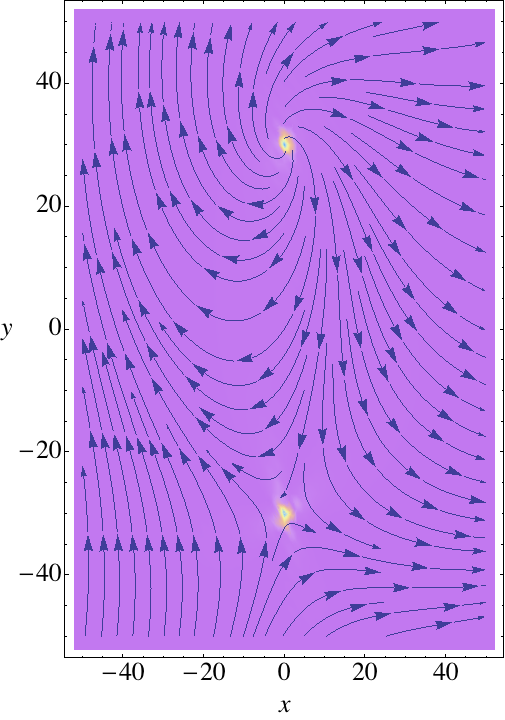} \includegraphics[width=0.4933\columnwidth]{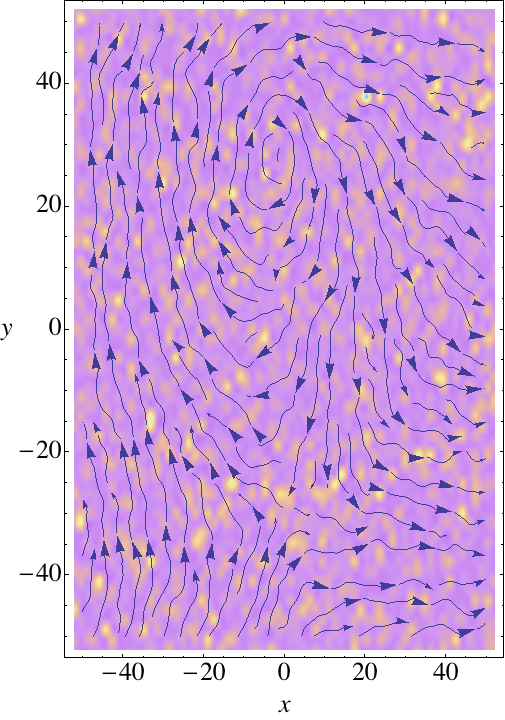}
\caption{\label{energy density} \textit{Left: }Polarization field $\bm{s}(\bm{r})$
and total energy distribution at $T=0$ for a vortex-antivortex pair
located at $(0,\pm30)$. Brighter regions characterize larger energy
density, where vortices or vortex-antivortex pairs are located. The
angle between the axis of the large vortex-antivortex pair and the
asymptotic uniform polarization is chosen to be $45^{\circ}$. \textit{Right:}
Configuration at temperatures close to $T_{c}$, where a large number
of thermally excited small vortex-antivortex pairs strongly alters
the power law of attraction in the given vortex-antivortex pair at
$(0,\pm30)$.}
\end{figure}

\begin{center}
\vspace*{-2em}

\par\end{center}

\section{Plasma Analogy} 

The ``potential'' $\Phi\left(\bm{r}\right)$ in (\ref{eq: Simplified VAP interaction})
is introduced in analogy with the electrostatics of charges: 
\begin{equation}
\Phi\left(\bm{r}\right)=\sum_{j}q_{j}F\left(\bm{r}-\bm{r}_{j}\right)=\int F\left(\boldsymbol{R}\right)\rho_T\left(\bm{r}^{\prime}\right)d^{2}\bm{r}^{\prime}.\label{eq: Expression for Phi}
\end{equation}
 Here, $\boldsymbol{R}=\bm{r}-\bm{r}^{\prime}$, $F\left(\boldsymbol{R}\right)=-2\pi\ln\left(1+\alpha R\right)-K_{0}R$,
and the vortex topological charge density $\rho_T\left(\bm{r}\right)=\sum_{j}q_{j}\delta^{\left(2\right)}\left(\bm{r}-\bm{r}_{j}\right)$.

By its definition, the potential
$\Phi\left(\bm{r}\right)$ satisfies a Poisson type equation, $\hat{L}_{\bm{r}}\Phi\left(\bm{r}\right)=\rho_{T}\left(\bm{r}\right)$.
Here the linear operator $\hat{L}_{\bm{r}}$ is defined such that
$\hat{L}_{\bm{r}}F\left(\bm{r}-\bm{r}^{\prime}\right)=\delta^{\left(2\right)}\left(\bm{r}-\bm{r}^{\prime}\right)$,
which in Fourier representation reads $L_{\bm{k}}=1/F_{\bm{k}}=\left\{ 2\pi K_{0}k^{-3}+4\pi^{2}\alpha/\left[k^{2}(\mathit{k+\alpha)}\right]\right\} ^{-1}$.
Let us follow the electrostatic analogy further. Since the energy
of a charge (vortex) $q$ placed in the external potential $\Phi$
is $U=q\Phi,$ the force, acting on the charge (vortex) is $\bm{F}=q\bm{E}$,
where the quasi-electric field vector $\bm{E}=-\nabla\Phi$, the mean field at the charge location. The energy of the pair in this
field is 
\begin{equation}
U=q_{+}\Phi\left(\bm{r}_{+}\right)+q_{-}\Phi\left(\bm{r}_{-}\right)\approx-\bm{d}\cdot{\bm{E}}\label{eq: energy of the pair}
\end{equation}
Here the topological dipole moment of a pair is introduced according
to $\bm{d}=\sum_{j}q_{j}\bm{r}_{j}=q_{+}\bm{r}_{+}+q_{-}\bm{r}_{-}\equiv\bm{r}$,
where $\bm{r}=\bm{r}_{+}-\bm{r}_{-}$. The density of the topological
polarization charges is 
\begin{equation}
\rho_{TP}=-\nabla\cdot\bm{P},\label{eq:Topological polarization charge density}
\end{equation}
where 
\begin{equation}
\bm{P}\left(\bm{r}\right)=n_{P}\left\langle \bm{d}\right\rangle \label{eq: Polarization vector}
\end{equation}
is the polarization vector of the vortex-antivortex pair gas, and $n_{P}$
is the surface density of vortex-antivortex pairs; $\left\langle ...\right\rangle $
denotes statistical averaging. After averaging the polarization topological
charge inside a contour $C$, $Q_{TP}=\int df\rho_{TP}$, becomes
a fractional number in general, while before averaging it must be
integer.

The equation for the potential of a point charge $Q$ placed at the
origin is $\Phi_{Q}\left(\bm{r}\right)$ is $\hat{L}_{\bm{r}}\Phi_{Q}\left(\bm{r}\right)=\rho_{T}$,
where $\rho_{T}=Q\delta^{\left(2\right)}\left(\bm{r}\right)+\rho_{TP}$.
In the weak-field approximation (whose validity we explain in detail in Appendix \ref{WFAproof}), that is
keeping only the linear terms in $\nabla\Phi$ of Eq.\,\eqref{eq: Most general form of WFA expression for polarization vector},
we obtain a general formula for $\bm{P}\left(\bm{r}\right)$: 
\begin{equation}
\bm{P}\left(\bm{r}\right)=-\int df^{\prime}\psi\left(\left|\bm{r}-\bm{r}^{\prime}\right|\right)\nabla\Phi\left(\bm{r}^{\prime}\right).\label{eq: Most general form of WFA expression for polarization vector}
\end{equation}
Here $\psi\left(\left|\bm{r}-\bm{r}^{\prime}\right|\right)$ is a
general nonlocal kernel in an isotropic and translationally invariant
medium, connecting electric field and polarization \cite{LL8}. Hence
$\left(\rho_{TP}\right)_{\bm{k}}=-k^{2}\psi\left(\boldsymbol{k}\right){\Phi_{Q\bm{k}}}$,
where $\psi\left(\boldsymbol{k}\right)=\int df$$\exp\left(-i\bm{k}\cdot\bm{r}\right)\psi\left(\boldsymbol{r}\right)$.
Our aim is to describe large-scale effects in a slowly varying field
$\Phi_{Q}\left(\bm{r}\right)$ corresponding to large $r$ and small
$k$. Henceforth, we can thus take $\left(\rho_{TP}\right)_{\bm{k}}\approx-\chi k^{2}\Phi_{Q\bm{k}}$,
where $\chi\equiv\psi\left(\boldsymbol{k}=0\right)=\int df\psi\left(\bm{r}\right)$.
This approximation holds when the kernel $\psi\left({\bm{r}}\right)$ decays fast
enough with $r$, such that $\int df\psi({\bm{r}})$ converges. The
physical reason behind the latter assumption is that small pairs are far
from dissociation and, thus, the vortex-antivortex pair gas
at any moment can be approximately subdivided into separated vortex-antivortex
pairs. In this approximation the polarization vector is given by (\ref{eq: Polarization vector}),
where the mean dipole moment of a pair $\left\langle \bm{d}\right\rangle $
is calculated with the help of a Boltzmann type formula with potential energy
(\ref{eq: energy of the pair}). It is clear from this qualitative
consideration that the characteristic length of $\psi\left({\bm{r}}\right)$
should be of order of the typical small pair dimension, $R_{P}$:
The mean size of a typical small
pair with unscreened interaction energy $u\left(r\right)$ is given by $R_{P}=J_{1}/J_{0}$.
Here, we defined thermal averages of moments of the radial distance 
as follows
\begin{align}
J_{n}&=\int d^{2}rr^{n}\exp\left[-u\left(r\right)/T\right]\nonumber\\
&=2\pi\intop_{0}^{\infty}drr^{n+1}e^{-\gamma r}\left(1+\alpha r\right)^{-\beta},\label{defJn}
\end{align} 
where $\beta=2\pi/T$, $\gamma=K_{0}/T$. 
At $T\approx T_{c}$, all parameters approach unity (in our units), 
therefore $R_{P}\sim1$ as well. 
Note that close to the phase transition this length scale
is comparable with the other typical scale of our problem, $\sim n_{P}^{-1/2}$,
therefore there is only one characteristic scale at short distances close to $T_c$.
So, for the slowly varying field 
\begin{equation}
\bm{P}\left(\bm{r}\right)\approx\chi\bm{E}\left(\bm{r}\right)=-\chi\nabla\Phi\left(\bm{r}\right).\label{eq: Polarization vector for slowly varying fields}
\end{equation}
Therefore $\chi$ is the {\em susceptibility of the vortex-antivortex
pair gas}. Putting everything together, we have for the potential
\begin{equation}
\Phi_{Q}(\bm{r})=Q\int\frac{d^{2}k}{\left(2\pi\right)^{2}}\frac{\exp\left(i\bm{k}\cdot\bm{r}\right)}{L_{\bm{k}}+\chi k^{2}}.\label{eq: Expression for F}
\end{equation}
We define the distance scale 
\begin{equation}
r_{0}=\frac{1}{2\pi K_{0}\chi},\label{eq: scale r_0}
\end{equation}
which involves the vortex-antivortex pair tension $K_{0}$ and the
susceptibility $\chi$ of the vortex-antivortex pair gas. At large
distances, $r\gg r_{0}$, the dominant contribution to the integral
comes from small values of $k$, for which the $L_{\bm{k}}$ term
in the denominator is negligible. Therefore, since $\left(\ln r\right)_{\bm{k}}=-2\pi/k^{2}$,
the potential of the {}``charge'' at large distances is logarithmic:
$\Phi_{Q}(\bm{r})=-\left(Q/2\pi\chi\right)\ln\left(r/C_{1}\right)$, 
where the constant $C_{1}\sim r_{0}$. In the opposite limit, $r\ll r_{0}$,  
the potential is linear: $\Phi_{Q}(\bm{r})\approx-QK_{0}r$.
We propose a simple interpolating expression: $\Phi_{Q}(\bm{r})\approx-\left(Q/2\pi\chi\right)\ln\left(1+r/r_{0}\right)$,
so that the energy of a sufficiently large pair of size $R$ is given
by 
\begin{equation}
U\left(R\right)=-\Phi_{Q=1}(R)=\frac{1}{2\pi\chi}\ln\left(1+\frac{R}{r_{0}}\right).\label{eq: Modified VAP energy}
\end{equation}
The distance scale $r_{0}$ in Eq.\,\eqref{eq: scale r_0} represents
an analogue of the Debye shielding radius for 2D interactions of topological
charges. We refer the reader for further details on the properties of the distance scale $r_0$ to the Appendix \ref{r0}.

\section{The Transition Temperature} 

The standard calculation procedure of the transition temperature for a gas of
polarizable vortex-antivortex pairs with interaction (\ref{eq: Modified VAP energy})
gives the following implicit equation for the transition temperature
in terms of the susceptibility \cite{Berezinskii,kosterlitz1973oma,kosterlitz1974critical},
$T_{c}=1/4\pi\chi$. At the transition temperature, $T=T_{c}$, vortex-antivortex
pairs begin to dissociate. This implies that at $T_{c}-T\ll T_{c}$
only a small fraction of the pairs is large and close to dissociation.
For this reason it is possible to neglect the interactions between
the largest vortex-antivortex pairs and calculate the energy of a
single large pair approaching its dissociation limit, which is permeated
by a cloud of comparatively small bound vortex-antivortex pairs. Hereinafter
we will subdivide vortex-antivortex pairs into two classes: small
pairs and large, close to dissociation, pairs. The shielding effect
arises due to the polarization cloud provided by small pairs, influencing
the potential energy of a large pair. 

Let us calculate first the polarizability
$\alpha_{P}$ of a single small pair. The energy of a small pair
in an external field $\bm{E}$ equals $V\left(\bm{r}\right)=u\left(r\right)-\bm{r\cdot E}$.
The average dipole moment of the small pair and the susceptibility
of the vortex-antivortex pairs gas within the framework of the weak-field
approximation (cf.\,Appendix \ref{WFAproof})  
are given by the relations $\left\langle \bm{d}\right\rangle =\int df\bm{r}\exp\left(-V/T\right)/\int df\exp\left(-V/T\right)\approx\alpha_{P}\bm{E},$
$\chi=\alpha_{P}n_{P}$. Here, $\alpha_{P}=J_{2}/\left(2TJ_{0}\right)$
is the small pair polarizability we are looking for, where we used the definition \eqref{defJn}. The potential
energy of a small pair in the field of the charge $Q$, $\left\langle \bm{d}\right\rangle \nabla\Phi\approx\left\langle \bm{d}\right\rangle \nabla\Phi_{V}$$\left(\boldsymbol{\rho}\right)$,
approximately does not depend on the position of the pair, $\boldsymbol{\rho}$,
so that $n_{P}$ can be considered as a constant.

\begin{figure}[t]
\includegraphics[width=0.85\columnwidth]{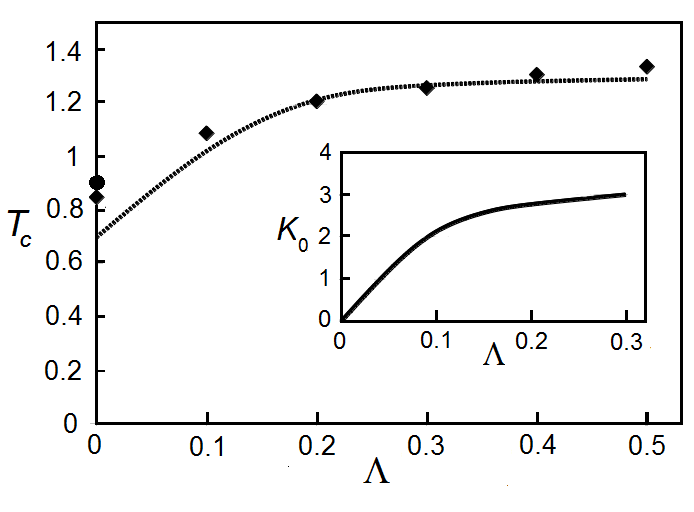} \caption{Transition temperature $T_{c}$ versus dipole-dipole interaction coupling
strength $\Lambda$ computed using Eq. (\ref{eq: New implicit equation for T_C})
(dashed line), and numerical result for $T_{c}$ (diamonds). The BKT
result \cite{Tobochnik,Ferrer}, corresponding to $\Lambda=0$, is
shown by a full circle on the $\Lambda=0$ axis. In the inset the
saturation effect for the vortex-antivortex pair tension coefficient
$K_{0}$ with increasing dipole-dipole interaction strength $\Lambda$
is displayed. The functional dependence $K_{0}=K_{0}(\Lambda)$ forms
the input for the evaluation in Eq.\,\eqref{eq: New implicit equation for T_C}
of the critical temperature $T_{c}$. \label{fig:Computed-transition-temperature}} 
\end{figure}

The surface density of vortex-antivortex pairs, $n_{P}$, at $T_{c}$
can be calculated as follows. At $T\approx T_{c}$ the vortex-antivortex
pairs only start to dissociate and the fraction of large pairs is
small. The typical pair is small, and the interaction $\approx u\left(r\right)$
between its vortices is still unscreened. 
Then, we conclude from basic arguments of statistical mechanics that the partition function of $N$ pairs on a surface with area $A$ is given by $z=z_{1}^{N}/N!\approx\left(z_{1}e/N\right)^{N}$, using
Stirling's formula, and 
where $z_{1}=\int d^{2}r_{1}d^{2}r_{2}\exp\left[-u\left(\left|\boldsymbol{r}_{1}-\boldsymbol{r}_{2}\right|\right)/T\right]=AJ_{0}$, 
which again employs \eqref{defJn}. 
Minimization of the free energy $F_{P}\left(N\right)=-T\ln z$
gives $N=z_{1}$, $n_{P}=N/A=J_{0}$. Using this result and the relation
$T_{c}=(4\pi\chi)^{-1}$, there follows $2\pi J_{2}=1$, leading to
an implicit equation for the critical temperature 
\begin{equation}
4\pi^{2}\intop_{0}^{\infty}dr\frac{r^{3}\exp\left(-K_{0}r/T_{c}\right)}{\left(1+\alpha r\right)^{2\pi/T_{c}}}=1.\label{eq: New implicit equation for T_C}
\end{equation}
Next we compare the semianalytical solution for $T_{c}$ from (\ref{eq: New implicit equation for T_C})
with the results of the numerical calculation, cf.\,Fig.\ref{fig:Computed-transition-temperature}.
As described in detail in Appendix \ref{Numerics}, we used Langevin Dynamics and
Binder's method to calculate $T_{c}$ in a series of simulations with
increasingly larger realizations of the model system. In the non-dipolar 
case, $K_{0}=0$ (or $\Lambda=0$), the equation (\ref{eq: New implicit equation for T_C})
yields $T_{c}\approx0.7$. From our numerical calculations, we found
that $T_{c}=0.85$. At large $\Lambda$ the transition temperature
tends to the constant value $T_{c}(\infty)\approx1.4$. From (\ref{eq: New implicit equation for T_C})
we derive $T_{c}\left(\infty\right)\approx1.3$. Note that the RG
arguments of \cite{MaierSchwabl} gave the much larger prediction
$T_{c}(\infty)=2\pi$. This in turn implies that it is rather difficult
to account for all essential Feynman graphs in order to adequately
describe the shielding effect in a RG calculation
(which in fact applies equally well to the short-range case $\Lambda=0$).

We note that the equation \eqref{eq: New implicit equation for T_C} is based on
the assumption of the noninteracting small pairs. Rigorously, though, as $R_P \sim n_P^{-1/2}$,  
this fails around $T\simeq T_c$. Hence we expect this approximation to give 
only an order of magnitude estimate for the susceptibility $\chi$, and the equality sign in 
\eqref{eq: New implicit equation for T_C} is to be replaced in that rigorous sense by an $\approx$. 
On the other hand, we find from our numerical calculations that the approximation leading to
 \eqref{eq: New implicit equation for T_C} still proves to be rather reliable for a sufficiently accurate 
prediction of the critical temperature, cf.\,Fig\,\ref{fig:Computed-transition-temperature}.

\section{Conclusion}
We have demonstrated both numerically and analytically
using an analogy to plasma physics, that vortex-antivortex pairs in
the 2D dipolar XY-model dissociate at a critical temperature in a
manner familiar from the BKT transition. This is due to the dipole-interaction-induced linear
confinement potential in an isolated large vortex-antivortex pair 
being shielded by a gas of small pairs, in which the large pair becomes
immersed around the transition point. Therefore, the logarithmic attraction
between vortices in large pairs is restored. By obtaining a physically
transparent scenario, we have therefore provided an unambiguous proof
that the BKT mechanism is applicable to a much broader class of systems
than hitherto established. Our simulations, combined with the analytical
approach presented above, give a rigorous confirmation of the qualitative
assumptions discussed in \cite{Physics of particle and nuclei letters}.
Shielding of the linear interaction in a vortex-antivortex pair implies
that the dipole interaction term $G_{dd}$ is irrelevant (in the sense of the RG), and that
the 2D dipolar XY-model belongs to the same universality class as
the contact interaction BKT-model corresponding to $G_{dd}=0$. This conclusion is
confirmed in Appendix \ref{Numerics} by arguments based on the numerical calculation
of the Binder cumulant. While the scale hierarchy, discussed
in our paper only qualitatively, can more adequately be formulated
in the language of the RG \cite{MaierSchwabl}, we have provided
clear evidence that a correct understanding of the physical nature of the phase transition 
can not be obtained within the RG approach.

We finally stress that the planar dipolar XY model (\ref{eq: Hamiltonian})
fundamentally differs from the commonly studied 2D system with all
dipoles oriented perpendicular to the plane \cite{scalar 1}, cf.,
e.g., \cite{Filinov}, which is a {\em scalar} model. Also, it
differs from a ``purely'' dipolar model (see \cite{baek2011},
for example), whose Hamiltonian does not contain the short-range (gradient) term, $G_{0}$,
in Eq.\,\eqref{eq: Hamiltonian}. Therefore, the physics behind these models is fundamentally different
from the dipolar XY case. For example, the ground state for square lattice
in \cite{baek2011} is antiferroelectric, but we have a ferroelectric ground state.

\acknowledgments 
The work of AYuV, AET, LIM, and POF was supported by Quantum Pharmaceuticals.
The research of URF was supported by the NRF of Korea, Grant Nos.
2010-0013103 and 2011-0029541.

\appendix

\section{On the microscopic derivation of the hydrodynamic Hamiltonian}
\label{microscopic}
We briefly outline in what follows the derivation of the Hamiltonian \eqref{eq: Hamiltonian},  
underlying our analysis, from microscopics.
The interaction energy of $N_M$ polar molecules with
dipole moments $\mu_A$, where $A = 1,\ldots,N_M$, equals 
\begin{equation}
V = \frac12 \sum_{\alpha\beta,A\neq B}\mu_{A\alpha}\mu_{B\beta}f_{\alpha\beta}(\rho) = V^L+V^S,
\end{equation}
where $\alpha,\beta$ are the spatial indices, ${\bm \rho}={\bm r}_A-{\bm r}_B$, and $f_{\alpha\beta}({\bm \rho})=
 f^L_{\alpha\beta}({\bm \rho})+ f^S_{\alpha\beta}({\bm \rho})$. 
 Here $f^L_{\alpha\beta}({\bm \rho})$
 describes the long-range interaction of molecules $f^L_{\alpha\beta}({\bm \rho}) = (\delta_{\alpha\beta}-3n_\alpha n_\beta)/\rho^3$,
 ${\bm n}={\bm \rho}/\rho$.   
The function $f_{\alpha\beta}^S({\bm \rho})$ represents the
short-range part, for example the
hydrogen bond interaction in the case of water
molecules. In the continuum approximation of hydrodynamics, 
when the polarization vector, ${\bm P}_{d} ({\bm r}) 
= \sum^{N_M}_{A=1} {\bm \mu}_A \delta^{(D)}({\bm r}-{\bm r}_A)$, 
is considered as a slowly
varying function, the latter replaces ${\bm \mu}_A$, 
and the summation over molecules $\sum_{A=1}^{N_M}$ 
is replaced by an integration $\int d^D r_A n_M$, where $n_M$ is the moment density and $D$ the spatial dimension. 
The hydrodynamic 
approximation is capable to describe the long-range
effects which are considered in our paper. After an integration by parts,  
the long-range part $V^L$ takes the form of \eqref{eq: dipole-dipole interaction}. 

In the short-range term, we can take $\rho\ll {\bm R}_{AB}=|{\bm R}_A +{\bm R}_B|/2$, and the energy density is expanded in $\rho$. 
Writing 
\begin{eqnarray}
V^{S}=\frac{1}{2}n_{M}^{2}\int d^{D}r_{A}d^{D}r_{B}P_{d,\alpha}\left({\bm r}_{A}\right)P_{d,\beta}\left({\bm r}_{B}\right)f_{\alpha\beta}^{S}\left({\bm \rho}\right),\nonumber\\
\end{eqnarray}
we use the following definitions and expansions
\begin{subequations}
\begin{align}
{\bm r}_{A} &=  {\bm R}_{AB}+\frac{1}{2}{\bm \rho},\; {\bm r}_{B}={\bm R}_{AB}-\frac{1}{2}{\bm \rho},\\
d^{D}r_{A}d^{D}r_{B}&=d^{D}R_{AB} d^{D}\rho,\; d^{D}\rho=\rho^{D-1}d\rho d\Omega_{n}, \\
P_{d,\alpha}\left({\bm r}_{A}\right)&\approx P_{d,\alpha}\left({\bm R}_{AB}\right)+\frac{1}{2}
\left({\bm \rho}\nabla_{R}\right)P_{\alpha}\left({\bm R}_{AB}\right),\\ 
P_{d,\beta}\left({\bm r}_{B}\right)&\approx  P_{d,\beta}\left({\bm R}_{AB}\right)-\frac{1}{2}\left({\bm \rho}\nabla_{R}\right)P_{\beta}\left({\bm R}_{AB}\right).
\end{align}
\end{subequations}
The tensor $f_{\alpha\beta}^{S}\left(\boldsymbol{\rho}\right)$ depends
only on the distance vector $\boldsymbol{\rho}$. Due to space isotropy it should
have the same form in any Cartesian frame, therefore
\begin{equation}
f_{\alpha\beta}^{S}\left(\boldsymbol{\rho}\right)=A\left(\rho\right)\delta_{\alpha\beta}+B\left(\rho\right)n_{\alpha}n_{\beta}.
\end{equation}
After integration over $d\Omega_{n}$, the nonvanishing contributions
contain even powers of $\boldsymbol{n}$ only.
Finally, identifying ${\bm s}  = {\bm P}_d/(\mu n_M)$ (assuming that all $\mu_A$ have the magnitude $\mu$),  
the integration 
$\intop_{0}^{\infty}\rho^{D-1}d\rho...$ reproduces the bilinear terms in \eqref{eq: G H}. 

The quartic ``self-interaction'' term, $\propto ({s}^2-1)^2$, stems from a spin saturation effect, $s\rightarrow 1$ for large moment densities
$n_M$. This saturation can be explained, for example, by the Langevin formula $s = L\left[\frac{\mu E_d}{k_B T}\right]$, with the Langevin function 
$L(x) = \coth (x) -1/x$, yielding $s\rightarrow 1$ for a large polarizing electric field $E_d$. 


\section{Physical meaning of the distance scale $r_{0}$}
\label{r0} 

The polarization topological charge density of the vortex-antivortex pair
gas close to a single topological charge $Q$ equals: $\rho_{TP}=-\nabla\cdot\bm{P}\approx-\chi QK_{0}/r$.
We conclude that the total charge inside a circle of radius $r$ is
given by $Q_{t}\left(r\right)=Q+Q_{TP}\approx Q\left(1-2\pi K_{0}\chi r\right)$.
From this qualitative consideration we thus come to an important conclusion:
the charge is essentially {\em shielded}, $Q_{t}\approx0$, which
occurs at a distance scale $r\sim r_{0}$. Close to the phase transition
$r_{0}\sim R_{P}\sim1$. The polarization of the vortex-antivortex
pair gas inhibits the linear attraction within large vortex-antivortex
pairs which would prevail with shielding not taken into account.
Hence we are led to conclude that the phase transition associated with the dissociation
of pairs is qualitatively very similar to the BKT transition in a
system with $\Lambda\rightarrow0$.

Starting from the order of magnitude estimate above, we now consider a more rigorous approach. According
to (\ref{eq: Simplified VAP interaction}) two unshielded, probe charges
at a distance $r$ interact as $G=-q_{1}q_{2}K_{0}r=q_{1}\Phi_{q_{2}}(r)$
(in the presented qualitative consideration we neglect the logarithmic
term in $u\left(r\right)$), where $\Phi_{q_{2}}(r)=-q_{2}K_{0}r$
is the topological potential of the charge $q_{2}$. Similarly, for
the point charge $Q$ placed at the origin and the probe charge $q$:
$G=q\Phi_{Q}(r)$, where $\Phi_{Q}\left(\bm{r}\right)\approx-Q_{t}\left(r\right)K_{0}r$.
From here and Eqs.\,(\ref{eq:Topological polarization charge density}),
(\ref{eq: Polarization vector for slowly varying fields}), we conclude 
\begin{equation}
\rho_{TP}\left(r\right)\approx-\chi K_{0}\triangle\left(rQ_{t}\right)=-\frac{\chi K_{0}}{r}\frac{d}{dr}\left[r\frac{d}{dr}\left(rQ_{t}\right)\right].
\end{equation} 
This yields a differential equation for $Q_{t}\left(r\right)$:
\begin{align}
Q_{t}\left(r\right)&=Q+\intop_{0}^{r}2\pi rdr\rho_{TP}\left(r\right)\nonumber\\
&=Q-2\pi\chi K_{0}r\frac{d}{dr}\left(rQ_{t}\right).
\end{align}
Its solution is given by 
\begin{equation}
Q_{t}\left(r\right)=\frac{Qr_{0}}{r}e^{r_{0}/r}\intop_{r_{0}/r}^{\infty}dx\frac{e^{-x}}{x}.
\end{equation} 
This total charge monotonically diminishes with $r$ from $Q_{t}=Q$
at $r=0$ to $Q=0$ at $r=\infty$. From the formula above for the effective charge shielding we again conclude that
its characteristic scale is $r_{0}$.

\section{Accuracy of the weak-field approximation}
\label{WFAproof}

We verify in this part of the Appendix the applicability of the 
weak-field approximation.
Using Eq.\,(\ref{eq: Modified VAP energy}), the average interaction
between the vortices at finite temperature is $\left\langle U\left(R\right)\right\rangle =z_{P}^{-1}\intop_{0}^{\infty}dR\, U\left(R\right)g\left(R\right)=T_{c}\left(1-4\tau\right)/\left[\left(-\tau\right)\left(1-2\tau\right)\right]$,
where $z_{P}=\intop_{0}^{\infty}dRg\left(R\right)$, $g\left(R\right)=R\exp\left[-U\left(R\right)/T\right]$,
and the relative temperature $\tau=\left(T-T_{c}\right)/T_{c}$. On 
the other hand, the typical size of a close to dissociation large pair,
$R_{DP}$, can be estimated from the relation $\left\langle U\left(R\right)\right\rangle \equiv\left(2\pi\chi\right)^{-1}\ln\left(1+R_{DP}/r_{0}\right)$.
Therefore, next to the phase transition, $\left|\tau\right|\ll1$,
the dimension of a dissociated pair is large, $R_{DP}\sim r_{0}\exp\left(1/\left|\tau\right|\right)$,
and its topological {}``electric'' field is small, $\boldsymbol{E}=-\nabla U\left(R\right)\propto\exp\left(-1/\left|\tau\right|\right)\ll1$.
A condition for the applicability of the weak-field approximation
is therefore the existence of a small parameter $\exp\left(-1/\left|\tau\right|\right)$,
representing the ratio of the typical topological {}``electric''
field of large pair, $\sim\left|\nabla U\left(R\right)\right|$ to
the field inside a small pair, $\sim\left|\nabla u\left(R\right)\right|$.

At $|\tau|\ll 1$, the dissociating pair
is so (exponentially) large, that most of the small
pairs inside the large pair are far away from the vortex sources
of the field such that the field created by
these sources is small. Therefore, the 
weak-field approximation is applicable close to the transition temperature.

\section{Numerics}
\label{Numerics}

We studied the thermodynamics of our model using Langevin Dynamics
\cite{schlick}. Using a discretized representation of the Hamiltonian
at every grid point $\alpha=1,2,...,N_{g}$, we performed a fixed
temperature run of sufficient length to get reliable averages. The
calculations were performed using periodic boundary conditions on
square lattices $L\times L$ with the number of independent nodes
$N_{g}=L^{2}$. The Langevin Dynamics dynamical equations are given
by: 
\begin{equation}
\frac{ds_{\alpha}}{dt}=-\gamma\frac{\partial}{\partial s_{\alpha}}(G_{0}+G_{dd})+\zeta_{\alpha},
\end{equation}
where $\gamma$ is a constant that determines the time scale of relaxation.
The stochastic thermal noise terms satisfy $\left\langle \zeta_{\alpha}(t)\right\rangle =0$
and $\left\langle \zeta_{\alpha}(t)\zeta_{\beta}(t')\right\rangle =2T\gamma\delta_{\alpha\beta}\delta(t-t')$.
In the Langevin Dynamics simulations we use a second-order Runge-Kutta
algorithm. The equations of motion above are integrated numerically
with the sufficiently small discrete time step $\triangle t=0.005$.
To compute the dipole-dipole interaction term in Eq.\,(2) in an efficient
$O(N_{g}\ln N_{g})$ way we used a NumPy FFTW realization \cite{oliphant2006guide}.

\begin{figure}[t]
 \includegraphics[width=0.95\columnwidth]{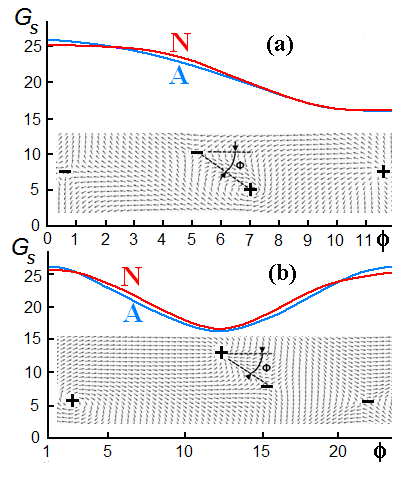} \caption{Free energy of a uniformly polarized 2D spin system with a small vortex-antivortex
pair in the center of a large pair (a) and for a slightly shifted
small pair (b) versus the angle (in units $\pi/12$) between the axis
of the small pair and the $x$ axis. Red (N) and blue (A) curves correspond
to the numerical calculation and the analytical approximation in Eq.\,(3),
respectively; $\Lambda=0.2$. \label{fig: Comparison for small in the center-1}}
\end{figure}

The software was first used to check the assumptions leading to Eq.\,\eqref{eq: Simplified VAP interaction}.
First we consider the case of zero temperature, $T=0$, and checked
the additivity rule expressed in (3) numerically by using an imaginary
time relaxation method, which is capable to find local minima of the
free energy functional on a configuration space of 2D vectors $\bm{s}_{\alpha}$
taken in nodes of a dense square lattice. Following \cite{belavin1975metastable},
the initial approximation was specified by the complex-valued skyrmion
type expression $s_{0}(z)=2W/\left(1+\left|W\right|^{2}\right)$,
with $\: W=\prod_{j=1}^{2N}\left(z-z_{j}\right)^{q_{j}},$ and $z=x+iy,$
$z_{j}=x_{j}+iy_{j}$. For a gas of pairs the total charge vanishes,
$\sum_{j}q_{j}=0$. Therefore, $W\rightarrow1$ at $z\rightarrow\infty$,
which agrees with the physical boundary condition $\boldsymbol{s}\rightarrow\left(1,0\right)$
at $r\rightarrow\infty$. At the first stage of imaginary time propagation
the cores of the vortex-antivortex pairs begin very slowly to approach
which, finally, leads to their annihilation. Due to this mutual attraction,
the vortex-antivortex pair therefore is not a real local minimum.
Any initial configuration inevitably transfers at $T=0$ to the absolute
ground state $\boldsymbol{s}\rightarrow\left(1,0\right)$. To stabilize
vortex-antivortex pairs, {}``pinning'' of vortices was used by adding
to the free energy a term $G_{{\rm pin}}=\sum_{j}G_{j},\: G_{j}=\int dfV_{j}\left(\bm{r}\right)\left[\bm{s}\left(\bm{r}\right)-\bm{s}_{0}\left(\bm{r}\right)\right]^{2}$,
where $V_{j}\left(\bm{r}\right)=V_{0}\exp\left[-\left(\bm{r}-\bm{r}_{j}\right)^{2}/a^{2}\right]$,
$a\sim1$. We investigated multiple configurations with different
numbers of vortex-antivortex pairs using the expression in Eq.\,\eqref{eq: Simplified VAP interaction},
and reproduced numerically the total energy within a small error,
less than $\sim5\%$ (examples of typical distributions are presented
in Fig.\,\ref{fig: Comparison for small in the center-1}). The string
tension constant $K_{0}$ was calculated by analyzing the mutual forces which prevail  
in a vortex-antivortex pair by calculating averages
of the derivatives of the pinning potential with respect to the positions
of the vortices. The accuracy we were able to achieve is limited by
the perturbations introduced by the pinning potential, and is sufficient
to prove the reliability of the additivity rule \eqref{eq: Simplified VAP interaction}, also cf.\,Fig.\,\ref{fig: Comparison for small in the center-1}.
We found that at $\Lambda\gtrsim\Lambda_{CR}\simeq0.4$, corresponding
to $K_{0}\left(\Lambda_{CR}\right)\approx3.1$, a physical instability
of the single vortex-antivortex pair configuration arises: A new small
vortex-antivortex pair is spontaneously created in the center of a
large vortex-antivortex pair. With further increase of $\Lambda$,
the {}``parent'' vortices are immersed into a cloud of small polarized
vortex-antivortex pairs, i.e. dipoles with zero total topological
charge. 
Polarization of vortex-antivortex pairs leads to a 
net topological charge density (cf. the discussion after \eqref{eq: Polarization vector}) and, hence, 
to a reduced increase of the line tension $K_{0}$ with $\Lambda$.
Ultimately, this leads to a saturation effect: $K_{0}\left(\Lambda\right)\approx K_{0}\left(\Lambda_{CR}\right)\approx3.1$
at $\Lambda\gtrsim\Lambda_{CR}$ (see the inset of Fig.\,\ref{fig:Computed-transition-temperature}).
The spontaneous creation of pairs follows also from the expression
(\ref{eq: Simplified VAP interaction}): At $\Lambda>\Lambda_{CR}$
the energy of the large ($R\gg1$) pair decreases with the emergence
of a new small pair in the center. This occurs at $\Lambda=\Lambda_{CR}$,
if we choose $\alpha=K_{0}\left(\Lambda_{CR}\right)/2\pi\approx K_{0}\left(\infty\right)$/$2\pi$.

\begin{figure}[b]
 \includegraphics[width=0.9\columnwidth]{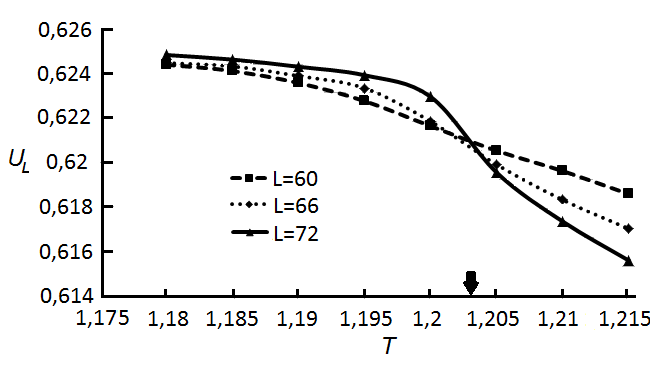} \caption{Explanation of Binder's approach to obtain the critical temperature
\cite{binder2010monte}. Binder's parameter $U_{L}$ was calculated
for $\Lambda=0.2$; the intersection point for different lattice sizes
$L$ yields $T_{c}$ (indicated by the bold arrow). \label{fig: Binder method}}
\end{figure}

To explore the critical behavior of the model depending on the dipole-dipole
interaction coupling constant $\Lambda$ numerically we use Binder's
method \cite{binder2010monte} and calculate the parameter $U_{L}=1-\left\langle \boldsymbol{s}^{4}\right\rangle /3/\left\langle \boldsymbol{s}^{2}\right\rangle ^{2}$
({}``Binder's cumulant'') versus temperature. Here $\boldsymbol{s}=\boldsymbol{S}/N_{g}$,
$\boldsymbol{S}=\sum_{\alpha=1}^{N_{g}}\boldsymbol{s}_{\alpha}$ and
the statistical averaging is done in a manner equivalent to the average
over polarization configurations obtained with Langevin Dynamics.
The intersection point of Binder's cumulants for different values
of the system size $L$ gives $T_{c}$, as shown in Fig.\,\ref{fig: Binder method}.
In the symmetric phase, $T>T_{c}$, $U_{L}=0+O\left(1/A\right)$ as
$A\rightarrow\infty$, where $A$ is the surface area. In the symmetry-broken
phase, $T<T_{c}$, $U_{L}=2/3+O\left(1/A\right)$. At the critical
point, $U_{L}$ tends towards a universal value $0<U_{L}^{\star}<2/3$,
which is specific for each model, and is determined by its universality
class \cite{binder2010monte,Binder Zeitschrift}. To determine the
universality class of the model at hand it seems natural to simply
compare our value $U_{L}^{\star}\approx0.621$ with that for the pure
BKT-model without dipole-dipole interaction. According to detailed
numerical results \cite{Selke,Baek2007}, the magnitude of $T_{c}$
does not depend on the simulation details, but, in fact, the value
$U_{L}^{\star}$ itself {}``...depends sensitively on boundary conditions,
details of the clusters used in calculating the cumulant, and symmetry
of the interactions or, here, lattice structure...,'' quoting Ref.\,\cite{Selke}.
As a consequence, we have to compare our result with others under
the same conditions. We know two such results for $\Lambda=0$: $U_{L}^{\star}\approx0.61$
for the XY-model \cite{Selke}, and $U_{L}^{\star}\approx0.62$ for
the generalized XY-model \cite{Qin2009}. As for the present dipolar
XY-model, one has $U_{L}^{\star}\approx0.62$ in a wide range of $\Lambda$
values \cite{rapini2007phase,mol2011phase}.


\begin{thebibliography}{References}
\bibitem{Linear behavior at small r} The function $u\left(r\right)$
contains logarithmic and linear terms at $r\gg1$ arising from $G_{0}$
and $G_{dd}$, respectively. At $r=0$ vortices annihilate, therefore
$u\left(0\right)=0$. At $r\ll1$ the linear behavior of $u\left(r\right)$
is imposed by two nearby topological charges, which stems from $G_{dd}$.
The chosen form of $u\left(r\right)$ accounts for these features.
\end{thebibliography}
\end{document}